\documentclass{aa}
\usepackage{psfig}
\usepackage{astrobib}
\usepackage{journals}
\begin{document}

   \thesaurus{01 (     
              	02.13.3;	
	      	11.01.2;  	
               	11.09.1 Mrk~348;
		11.10.1;	
               	11.19.1;	
		13.19.1	  	
)}

   \title{Discovery of a very luminous megamaser during a radio flare in
   the Seyfert 2 galaxy Mrk~348}

\authorrunning{Falcke et al.}
\titlerunning{Discovery of a megamaser in Mrk~348}


   \author{Heino Falcke\inst{1}, Christian Henkel\inst{1}, Alison
   B.~Peck\inst{1}, Yoshiaki Hagiwara\inst{1}, M. Almudena
   Prieto\inst{2}, Jack F. Gallimore\inst{3}}

   \offprints{hfalcke@mpifr-bonn.mpg.de}

   \institute{Max-Planck-Institut f\"ur Radioastronomie,
              Auf dem H\"ugel 69, 53121 Bonn, Germany \and
European Southern Observatory,
Karl-Schwarzschild-Stra\ss{}e 2,
85740 Garching, Germany \and
National Radio Astronomy Observatory,
520 Edgemont Rd.,
Charlottesville, VA 22903-2475, USA}

   \date{Astronomy \& Astrophysics Letters (2000), in press}

   \maketitle

   \begin{abstract} We report the detection of a new H$_2$O megamaser
   in the Seyfert 2 galaxy Mrk~348 with the MPIfR 100 m telescope in
   Effelsberg. With an apparent isotropic luminosity of $L_{\rm
   H_2O}\simeq420L_\odot$ the maser is the third most luminous maser
   discovered so far. The detected line is unusually broad ($\Delta
   v\simeq130$ km s$^{-1}$), is redshifted by $\sim130$ km s$^{-1}$
   from the systemic velocity, and is asymmetric with a pronounced
   blue wing. While evidence for obscuring material towards the
   nucleus of this galaxy was found earlier, this detection is the
   first direct observation of molecular material in the vicinity of
   the AGN. We also searched for absorption from ammonia (NH$_3$) and
   cyclopropenylidene (C$_3$H$_2$) against the bright radio
   nucleus. The H$_2$O line was only marginally detected in archival
   data indicating that the maser flared recently in conjunction with
   a huge radio continuum flare. Continuum and line flux density
   increased by a factor of three, suggesting an unsaturated maser.
   The radio continuum flare has made Mrk~348 the most radio luminous
   megamaser galaxy known. It is pointed out that megamaser galaxies
   contain a rather large fraction of galaxies with prominent radio
   cores and it is speculated that the flare in the maser emission in
   Mrk~348 is related to the flare in the nuclear jet.

       \end{abstract}

%

\section{Introduction}
Emission from H$_2$O masers has been found in a few galaxies,
exhibiting apparent isotropic luminosities a million times higher than
in typical stellar masers
\cite{DosSantosLepine1979,GardnerWhiteoak1982,ClaussenHeiligmanLo1984,HenkelGuestenWilson1984,HaschickBaan1985,BraatzWilsonHenkel1994}.
The detection rate of these megamasers is very low, i.e.~about 5\%
among Seyfert galaxies \cite{BraatzWilsonHenkel1997} and almost zero
among radio galaxies (e.g.,~\citeNP{HenkelWangFalcke1998}). The maser
is associated with dense and warm material, possibly a molecular torus
or disk, around an active galactic nucleus (AGN). The AGN apparently
produces the seed radio photons and the X-ray photons needed to pump
the masing material \cite{NeufeldMaloneyConger1994}.

With the help of Very Long Baseline Interferometry (VLBI) megamasers
can be used to investigate the small-scale structure of an AGN in
great detail. In the case of NGC~4258 this has helped to establish the
presence of a thin, warped disk around the nucleus, to determine the
black hole mass, and even to measure the precise distance to this
galaxy
\cite{MiyoshiMoranHerrnstein1995,HerrnsteinMoranGreenhill1999}. 

Finding new megamaser galaxies is therefore of prime interest. The
only clear trend that has emerged in recent years is that megamasers
are exclusively found in type 2 AGN, i.e.~those Seyferts and LINERs
which are expected to be obscured by a molecular torus according to
the unified scheme \cite{Antonucci1993}. Many have high absorbing
column depths inferred from X-ray spectroscopy. There is also an
indication of an excess of megamasers in highly inclined galaxies
\cite{BraatzWilsonHenkel1997,FalckeWilsonHenkel2000}. Here we report 
the discovery of a hitherto undetected and very luminous megamaser in
the Seyfert galaxy Mrk~348 during a radio flare of the AGN.

Mrk~348 (NGC~262, $z=0.01503$
\citeNP{HuchraVogeleyGeller1999}, luminosity distance $D=62.5$ Mpc for
$z$ converted into the Galactic Standard of Rest and $H_{0}=75$ km
s$^{-1}$ Mpc$^{-1}$), is a well-studied Seyfert 2 galaxy with broad
emission-lines in polarized light \cite{MillerGoodrich1990}. The
galaxy is classified as an S0 with a rather low inclination
($i=16^\circ$, see
\citeNP{BraatzWilsonHenkel1997}).  Ground-based
\cite{SimpsonMulchaeyWilson1996} and Hubble Space Telescope 
imaging \cite{FalckeWilsonSimpson1998} have revealed a dust lane
crossing the nucleus and an excitation cone in emission-lines. Ginga
observations found hard X-ray emission and a high absorbing column
depth of $N_{\rm H}=10^{23.1}\,{\rm cm}^{-2}$ towards the nucleus
\cite{WarwickKoyamaInoue1989}. All this suggests the presence of an
obscuring torus in Mrk~348. Attempts to detect the obscuring
material through radio spectroscopy have failed so far (e.g.~H I,
\citeNP{GallimoreBaumO'Dea1999}).

What makes this galaxy stand out among Seyfert galaxies is its bright
and variable radio nucleus. \citeN{NeffdeBruyn1983} found a compact
radio core on VLBI scales with a flux density of several hundred
milli-Jansky and a flat to inverted
spectrum. \citeN{UlvestadWrobelRoy1999} presented more recent VLBI
observations for two epochs, showing a two-component structure
expanding with sub-relativistic speeds. They also noted a flare of the
radio continuum emission at 15 GHz with the flux rising from 120 mJy
to 570 mJy between 1997.10 and 1998.75. In the following we will
present and discuss results of K-band radio spectroscopy of this
galaxy.

\section{Observations and Results}
Data were taken in March and April 2000, using the Effelsberg 100 m
telescope of the MPIfR equipped with a dual channel K-band HEMT
receiver. The system temperature was of order 70\,K on a $T_{\rm
A}^{*}$ temperature scale; the beam size was 40$''$. The data were
recorded using an autocorrelator with 8 $\times$ 256 channels and
bandwidths of 80\,MHz each. The eight backends were configured in two
groups of four, sampling the two orthogonal linear
polarizations. Frequency shifts between the four backends representing
a given linear polarization were adjusted in such a way that a total
velocity range of 3000\,km\,s$^{-1}$ could be covered.

The measurements were carried out in a dual beam switching mode
(switching frequency 1\,Hz) with a beam throw of 121$''$ in
azimuth. Only linear baselines were subtracted. Calibration was
obtained by measurements of W3OH (3.2\,Jy according to
\citeNP{MauersbergerWilsonHenkel1988}). Pointing could be checked on
Mrk~348 itself and was found to be stable to within 5--7$''$.

The galaxy was initially observed to look for ammonia (NH$_3$) and
cyclopropenylidene (C$_3$H$_2$) absorption against the bright radio
nucleus. No absorption features were found. The transitions, rest
frequencies, and upper limits of these observations are listed in
Table \ref{abslines}.

\begin{table}
\begin{center}
\begin{tabular}{cccc}
Line & $\nu$ & $\Delta v$ & $S_\nu$ \\
     & [GHz] & [km s$^{-1}$] & [mJy] \\ 
\hline
\hline
NH$_3$(1,1) & 23.694496 & 1 & $<17$ \\
NH$_3$(2,2) & 23.722631 & 1 & $<15$ \\
NH$_3$(3,3) & 23.870130 & 1 & $<11$ \\
NH$_3$(4,4) & 24.139417 & 1 & $<12$ \\
C$_3$H$_2$ 1$_{10}$--1$_{01}$ & 18.343146 & 5 & $<6$ \\
\end{tabular}
\end{center}
\caption[]{\label{abslines}
Upper limits for absorption lines measured towards the
nucleus of Mrk~348. 
Col. (1) -- molecule and transition observed;
Col. (2) -- rest frequency of transition in GHz;
Col. (3) -- channel width in km s$^{-1}$;
Col. (4) -- upper limits (1$\sigma$) for flux per channel in mJy.}
\end{table}

\begin{figure}
\centerline{\psfig{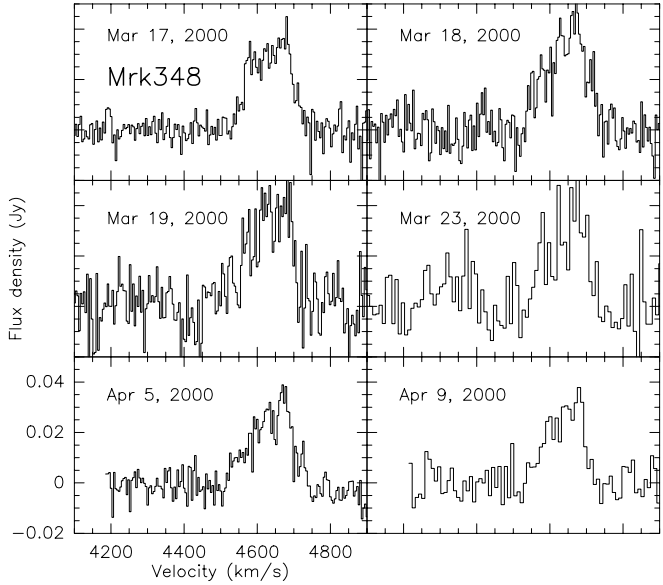}}
\caption[]{\label{h2ospec}Spectra of Mrk~348 ($\alpha$=00$^{\rm h}$
48$^{\rm m}$ 47.1$^{\rm s}$, $\delta$=31$^\circ$ 57$^\prime$
25$^{\prime\prime}$, J2000) taken with the 100 m telescope at
Effelsberg. The x-axis displays Local Standard of Rest (LSR)
velocities ($v_{\rm LSR}-V_{\rm Hel}=1.7$ km s$^{-1}$, $v_{\rm
sys}$=4509 km s$^{-1}$ ).  Velocities and channel widths follow the
optical convention. The broad H$_2$O maser line is clearly detected on
all six days at $v$=4640 km s$^{-1}$.}
\end{figure}

\begin{figure}
\centerline{\psfig{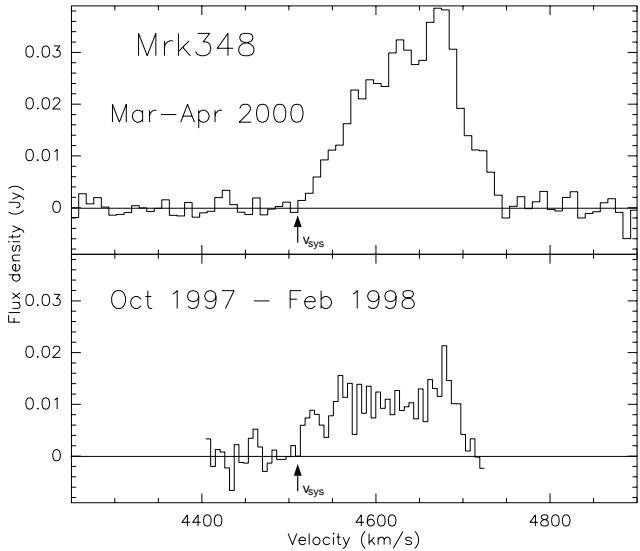}}
\caption[]{\label{specall}Combined spectra of Mrk~348 for spectra
taken in March -- April 2000 (top) and October 1997 -- February 1998
(bottom) indicating that the maser has flared significantly between
the two epochs. The arrow marks the systemic velocity.}
\end{figure}

Since the H$_2$O ($6_{16}-5_{23}$) transition is very close we also
observed the redshifted 22.23508 GHz line and detected an emission
feature on March 17. We repeated the observations on five subsequent
days and detected the line each time. The six spectra are shown in
Fig.~\ref{h2ospec}. We also produced a combined spectrum of all days
(Fig.~\ref{specall}, top) and fitted the H$_2$O line with a Gaussian
profile. A one component fit yields a central velocity of
$v=4641.6\pm1.2$ km s$^{-1}$ for the line, which is redshifted by 133
km s$^{-1}$ from the systemic velocity. The peak flux is $S_{\nu}=34$
mJy with a full width at half maximum of $\Delta v=130\pm3$ km
s$^{-1}$. The integrated flux is 4.71$\pm0.1$ Jy km s$^{-1}$, yielding
an apparent isotropic luminosity of 420 $L_\odot$. A two component fit
yields a narrow line at $v=4677.7\pm0.8$ km s$^{-1}$ with $\Delta
v=58.6\pm3.2$ km s$^{-1}$ and $S_{\nu}=27$ mJy plus a broad line with
$v=4609.5\pm1.9$ km s$^{-1}$, $\Delta v=101.8\pm1.9$ km s$^{-1}$, and
$S_{\nu}=26$ mJy. The latter fit indicates that the line is asymmetric
and has a pronounced blue wing.  If we compare our two best spectra
from 17 March and 5 April we tentatively find some variability in the
shape of this blue wing.  We searched for additional high velocity
features and did not detect anything down to a limit of 6 mJy
(1$\sigma$, 4 km s$^{-1}$) in the range 3250 to 6200 km s$^{-1}$ and
down to a limit of 10-15 mJy in the range $-1470$ to 10480 km
s$^{-1}$.  The continuum flux density we find at 22 GHz is $0.8\pm0.1$
Jy, corresponding to $4\cdot10^{23}$ Watt Hz$^{-1}$.

We also reduced some archival data of earlier observations of Mrk~348
taken between October 1997 and February 1998 and combined them into
one data set (Fig.~\ref{specall}, bottom). In this spectrum the broad
line is marginally detected with a flux density approximately three
times lower than in the current observations. The smaller and variable
baseline in these earlier observations made a reliable identification
of this line impossible without {\it a priori} information.

\section{Discussion and Summary}
We have clearly discovered a new megamaser in Mrk~348.  Its luminosity
is comparable to the emission from NGC~3079 which contains the second
most luminous H$_2$O maser after TXS~2226-184
\cite{KoekemoerHenkelGreenhill1995}. The line width is among the
broadest found for a megamaser, similar to the masers in TXS~2226-184
and NGC~1052. Assuming the emission is associated with material close
to the center this is the first spectroscopic evidence for molecular
gas possibly obscuring the nucleus.

Despite the bright radio continuum and the molecular maser line we
have found no absorption lines from either NH$_3$ or C$_3$H$_2$. This
is in line with the earlier non-detection of H {\sc I} absorption
\cite{GallimoreBaumO'Dea1999}. C$_3$H$_2$ is an
organic ring molecule which is widespread in the Galaxy and is
associated with diffuse gas in the ISM
\cite{MatthewsIrvine1985}. It was also found in absorption against
the nucleus of the radio galaxy Centaurus A
\cite{SeaquistBell1986}. From our non-detection we find a 1$\sigma$
upper limit for the optical depth $\tau$ times the covering factor $f$
of $\tau\cdot f<0.0075$. This is seven times smaller than the value
found for Centaurus A and might be related to the fact that we see
Mrk~348 almost face on.

On the other hand, with its type 2 AGN, polarized broad
emission-lines, and a nuclear dust lane seen, Mrk~348 falls right into
the roster of typical megamaser galaxies, where it is suggested that
the masing material is part of the obscuring `torus' in the unified
scheme of AGN. The face-on orientation of Mrk~348 then would suggest
that the axis of this torus and the galaxy disk axis are severely
misaligned.

In October 1997 -- February 1998 the flux density of the line was
three times lower than it was in March 2000. In the earlier survey by
\citeN{BraatzWilsonHenkel1996} the maser line was not detected and
given its line width and low flux density it would have resulted only
in a broad feature at the $3\sigma$ level. Linear interpolation with
time of the continuum flux density given in
\citeN{UlvestadWrobelRoy1999} suggests a continuum flux density for
Mrk~348 around 310 mJy in October 1997 at 15 GHz. The level we measure
is roughly a factor of three higher -- assuming a flat spectrum -- and
this increase in flux density is similar to the increase in flux
density of the line. This could indicate a correlation between
continuum and maser flux density, implying an unsaturated maser. With
its current radio luminosity the galaxy is now the most radio luminous
megamaser galaxy ever discovered. The response of the line to the
continuum flare within about 2 years sets an upper limit to the
distance of the masers from the nucleus of $\la$0.6 pc, which is of
similar order as the size scale of the molecular disk found in
NGC~4258.

Indeed, \citeN{BraatzWilsonHenkel1996} noted a certain excess of
detected megamaser galaxies with large radio powers. Mrk~348 certainly
adds to this trend. The distribution of radio powers at 6 cm of the
parent samples of AGN selected by
\citeN{BraatzWilsonHenkel1996} has  
a peak around $10^{21.75}$ Watt/Hz. If we add Mrk~348 and more
recently discovered megamasers
\cite{GreenhillHerrnsteinMoran1997,HagiwaraKohnoKawabe1997} and also
complement the radio data in \citeN{BraatzWilsonHenkel1996} with more
recent data from the NASA Extragalactic Database, we find that 12 out
of 19 detected megamaser galaxies are at or above the peak in the
distribution of radio power for galaxies without megamaser detections.

Because of the possibly biased selection of the detected megamasers
this is not highly significant. However, it highlights an apparently
necessary prerequisite for megamaser emission, namely an AGN with a
compact radio core to provide seed photons. So far all detected
megamasers have compact ($<1\arcsec$), mostly flat-spectrum, radio
emission at a level of a few milli-Jansky. In some cases, like Mrk~348
\cite{NeffdeBruyn1983}, NGC~1052 \cite{ShafferMarscher1979},
Mrk~1210
\cite{HeislerNorrisJauncey1998}, NGC~2639
\cite{WilsonRoyUlvestad1998}, NGC~3079 \cite{TrotterGreenhillMoran1998},
NGC~5793 \cite{WhiteoakGardner1987,GardnerWhiteoakNorris1992},
NGC~5506 \cite{SadlerSleeReynolds1995}, and possibly NGC~4945
\cite{ElmouttieHaynesJones1997} the radio cores can even reach several
tens to hundreds of milli-Jansky.

While compact radio cores in Seyfert and LINER galaxies are not
uncommon, only very few are so prominent as those in some of the
radio-bright megamaser sources. We find that all megamasers mentioned above,
i.e.~more than a third of known megamasers, have compact radio cores
above a fiducial limit of 25 mJy at 5 GHz. On the other hand, in a
survey of spiral galaxies
\citeN{SadlerSleeReynolds1995} find only 3 out of 54 galaxies (22 of
which are Seyfert galaxies) with compact cores above 25 mJy at 5
GHz. Similarly, in a survey of nearby AGN
\citeN{NagarFalckeWilson2000} find roughly 40\% of Seyfert and LINER
galaxies to contain compact flat-spectrum radio cores. However, only
three out of 48 galaxies have flux densities $>$25 mJy.  Known
megamasers therefore seem to prefer galaxies with relatively bright
compact radio emission.

Mrk~348 currently has the brightest and most prominent radio core
among megamaser galaxies.  Morphology, spectrum, and variability of
the core are very similar to the radio core in III~Zw~2 which was the
first Seyfert galaxy discovered to contain a superluminal jet. This
galaxy has a millimeter-peaked spectrum and a jet which shows a
stop-and-go behavior indicative of a strong interaction with dense
material on the sub-parsec scale
\cite{FalckeBowerLobanov1999,BrunthalerFalckeBower2000}. 
\citeN{UlvestadWrobelRoy1999}
therefore speculate whether the bright inverted radio core in Mrk~348
could be interpreted similarly to those in Compact Symmetric Objects
(CSOs) with a Gigahertz-Peaked-Spectrum (GPS, see
\citeNP{O'Dea1998}). In these galaxies bright hotspots are formed in a
jet that terminates already on the parsec scale. In III~Zw~2 and
Mrk~348 this seems to happen on even smaller scales, leading to higher
peak frequencies and could be due to frustration of the jet by a
molecular cloud or even a warped or misaligned torus.

Since the masers in NGC~1052, which have similar broad line widths as
in Mrk~348, are found along the radio jet
\cite{ClaussenDiamondBraatz1998} it should be checked whether in Mrk~348
one has an analogous situation. One can speculate that in such a case
the evolution of the radio flare and the evolution of the maser flare
and its blue wing could be related, possibly providing a unique
diagnostic tool to study jet-ISM interactions.

In any case, with its bright radio core Mrk~348 provides an ideal
opportunity to observe the maser lines in this galaxy at high
resolution with VLBI during this flare even though the lines
still have a rather low flux. Since radio and maser emission seem
to be highly variable both should be monitored frequently. Given that
Mrk~348 was not discovered in an earlier survey this finding also
suggests that existing samples should be revisited to search for more
flaring megamasers.

\begin{acknowledgements}
We thank Alan Roy for helpful discussions. We are grateful to Jim
Ulvestad for a prompt referee report and useful suggestions. This
research has made use of the NASA/IPAC Extragalactic Database (NED)
which is operated by JPL, Caltech, under contract with NASA.
\end{acknowledgements}

\clearpage


\begin{thebibliography}{}

\bibitem[\protect\citeauthoryear{{Antonucci}}{{Antonucci}}{1993}]{Antonucci199%
3}
{Antonucci} R., 1993, \araa, 31, 473

\bibitem[\protect\citeauthoryear{{Braatz}, {Wilson}, \& {Henkel}}{{Braatz}
  et~al.}{1994}]{BraatzWilsonHenkel1994}
{Braatz} J.~A., {Wilson} A.~S.,  {Henkel} C., 1994, \apjl, 437, L99

\bibitem[\protect\citeauthoryear{{Braatz}, {Wilson}, \& {Henkel}}{{Braatz}
  et~al.}{1996}]{BraatzWilsonHenkel1996}
{Braatz} J.~A., {Wilson} A.~S.,  {Henkel} C., 1996, \apjs, 106, 51

\bibitem[\protect\citeauthoryear{{Braatz}, {Wilson}, \& {Henkel}}{{Braatz}
  et~al.}{1997}]{BraatzWilsonHenkel1997}
{Braatz} J.~A., {Wilson} A.~S.,  {Henkel} C., 1997, \apjs, 110, 321

\bibitem[\protect\citeauthoryear{Brunthaler et~al.}{Brunthaler
  et~al.}{2000}]{BrunthalerFalckeBower2000}
Brunthaler A., Falcke H., Bower G.~C., et~al., 2000, \aap, Letters, in press

\bibitem[\protect\citeauthoryear{{Claussen} et~al.}{{Claussen}
  et~al.}{1998}]{ClaussenDiamondBraatz1998}
{Claussen} M.~J., {Diamond} P.~J., {Braatz} J.~A., {Wilson} A.~S.,  {Henkel}
  C., 1998, \apjl, 500, L129

\bibitem[\protect\citeauthoryear{{Claussen}, {Heiligman}, \& {Lo}}{{Claussen}
  et~al.}{1984}]{ClaussenHeiligmanLo1984}
{Claussen} M.~J., {Heiligman} G.~M.,  {Lo} K.~Y., 1984, \nat, 310, 298

\bibitem[\protect\citeauthoryear{{Dos Santos} \& {Lepine}}{{Dos Santos} \&
  {Lepine}}{1979}]{DosSantosLepine1979}
{Dos Santos} P.~M.,  {Lepine} J.~R.~D., 1979, \nat, 278, 34

\bibitem[\protect\citeauthoryear{{Elmouttie} et~al.}{{Elmouttie}
  et~al.}{1997}]{ElmouttieHaynesJones1997}
{Elmouttie} M., {Haynes} R.~F., {Jones} K.~L., et~al., 1997, \mnras, 284, 830

\bibitem[\protect\citeauthoryear{{Falcke} et~al.}{{Falcke}
  et~al.}{1999}]{FalckeBowerLobanov1999}
{Falcke} H., {Bower} G.~C., {Lobanov} A.~P., et~al., 1999, \apjl, 514, L17

\bibitem[\protect\citeauthoryear{{Falcke} et~al.}{{Falcke}
  et~al.}{2000}]{FalckeWilsonHenkel2000}
{Falcke} H., {Wilson} A.~S., {Henkel} C., {Brunthaler} A.,  {Braatz} J.~A.,
  2000, \apjl, 530, L13

\bibitem[\protect\citeauthoryear{{Falcke}, {Wilson}, \& {Simpson}}{{Falcke}
  et~al.}{1998}]{FalckeWilsonSimpson1998}
{Falcke} H., {Wilson} A.~S.,  {Simpson} C., 1998, \apj, 502, 199

\bibitem[\protect\citeauthoryear{{Gallimore} et~al.}{{Gallimore}
  et~al.}{1999}]{GallimoreBaumO'Dea1999}
{Gallimore} J.~F., {Baum} S.~A., {O'Dea} C.~P., {Pedlar} A.,  {Brinks} E.,
  1999, \apj, 524, 684

\bibitem[\protect\citeauthoryear{{Gardner} \& {Whiteoak}}{{Gardner} \&
  {Whiteoak}}{1982}]{GardnerWhiteoak1982}
{Gardner} F.~F.,  {Whiteoak} J.~B., 1982, \mnras, 201, 13P

\bibitem[\protect\citeauthoryear{{Gardner} et~al.}{{Gardner}
  et~al.}{1992}]{GardnerWhiteoakNorris1992}
{Gardner} F.~F., {Whiteoak} J.~B., {Norris} R.~P.,  {Diamond} P.~J., 1992,
  \mnras, 258, 296

\bibitem[\protect\citeauthoryear{{Greenhill} et~al.}{{Greenhill}
  et~al.}{1997}]{GreenhillHerrnsteinMoran1997}
{Greenhill} L.~J., {Herrnstein} J.~R., {Moran} J.~M., {Menten} K.~M.,
  {Velusamy} T., 1997, \apjl, 486, L15

\bibitem[\protect\citeauthoryear{{Hagiwara} et~al.}{{Hagiwara}
  et~al.}{1997}]{HagiwaraKohnoKawabe1997}
{Hagiwara} Y., {Kohno} K., {Kawabe} R.,  {Nakai} N., 1997, \pasj, 49, 171

\bibitem[\protect\citeauthoryear{{Haschick} \& {Baan}}{{Haschick} \&
  {Baan}}{1985}]{HaschickBaan1985}
{Haschick} A.~D.,  {Baan} W.~A., 1985, \nat, 314, 144

\bibitem[\protect\citeauthoryear{{Heisler} et~al.}{{Heisler}
  et~al.}{1998}]{HeislerNorrisJauncey1998}
{Heisler} C.~A., {Norris} R.~P., {Jauncey} D.~L., {Reynolds} J.~E.,  {King}
  E.~A., 1998, \mnras, 300, 1111

\bibitem[\protect\citeauthoryear{{Henkel} et~al.}{{Henkel}
  et~al.}{1984}]{HenkelGuestenWilson1984}
{Henkel} C., {G\"usten} R., {Downes} D., et~al., 1984, \aap, 141, L1

\bibitem[\protect\citeauthoryear{{Henkel} et~al.}{{Henkel}
  et~al.}{1998}]{HenkelWangFalcke1998}
{Henkel} C., {Wang} Y.~P., {Falcke} H., {Wilson} A.~S.,  {Braatz} J.~A., 1998,
  \aap, 335, 463

\bibitem[\protect\citeauthoryear{{Herrnstein} et~al.}{{Herrnstein}
  et~al.}{1999}]{HerrnsteinMoranGreenhill1999}
{Herrnstein} J.~R., {Moran} J.~M., {Greenhill} L.~J., et~al., 1999, \nat, 400,
  539

\bibitem[\protect\citeauthoryear{{Huchra}, {Vogeley}, \& {Geller}}{{Huchra}
  et~al.}{1999}]{HuchraVogeleyGeller1999}
{Huchra} J.~P., {Vogeley} M.~S.,  {Geller} M.~J., 1999, \apjs, 121, 287

\bibitem[\protect\citeauthoryear{{Koekemoer} et~al.}{{Koekemoer}
  et~al.}{1995}]{KoekemoerHenkelGreenhill1995}
{Koekemoer} A.~M., {Henkel} C., {Greenhill} L.~J., et~al., 1995, \nat, 378, 697

\bibitem[\protect\citeauthoryear{{Matthews} \& {Irvine}}{{Matthews} \&
  {Irvine}}{1985}]{MatthewsIrvine1985}
{Matthews} H.~E.,  {Irvine} W.~M., 1985, \apjl, 298, L61

\bibitem[\protect\citeauthoryear{{Mauersberger}, {Wilson}, \&
  {Henkel}}{{Mauersberger} et~al.}{1988}]{MauersbergerWilsonHenkel1988}
{Mauersberger} R., {Wilson} T.~L.,  {Henkel} C., 1988, \aap, 201, 123

\bibitem[\protect\citeauthoryear{{Miller} \& {Goodrich}}{{Miller} \&
  {Goodrich}}{1990}]{MillerGoodrich1990}
{Miller} J.~S.,  {Goodrich} R.~W., 1990, \apj, 355, 456

\bibitem[\protect\citeauthoryear{{Miyoshi} et~al.}{{Miyoshi}
  et~al.}{1995}]{MiyoshiMoranHerrnstein1995}
{Miyoshi} M., {Moran} J., {Herrnstein} J., et~al., 1995, \nat, 373, 127

\bibitem[\protect\citeauthoryear{{Nagar} et~al.}{{Nagar}
  et~al.}{2000}]{NagarFalckeWilson2000}
{Nagar} N.~M., {Falcke} H., {Wilson} A.~S.,  {Ho} L.~C., 2000, \apjl, submitted

\bibitem[\protect\citeauthoryear{{Neff} \& {de Bruyn}}{{Neff} \& {de
  Bruyn}}{1983}]{NeffdeBruyn1983}
{Neff} S.~G.,  {de Bruyn} A.~G., 1983, \aap, 128, 318

\bibitem[\protect\citeauthoryear{{Neufeld}, {Maloney}, \& {Conger}}{{Neufeld}
  et~al.}{1994}]{NeufeldMaloneyConger1994}
{Neufeld} D.~A., {Maloney} P.~R.,  {Conger} S., 1994, \apjl, 436, L127

\bibitem[\protect\citeauthoryear{{O'Dea}}{{O'Dea}}{1998}]{O'Dea1998}
{O'Dea} C.~P., 1998, \pasp, 110, 493

\bibitem[\protect\citeauthoryear{{Sadler} et~al.}{{Sadler}
  et~al.}{1995}]{SadlerSleeReynolds1995}
{Sadler} E.~M., {Slee} O.~B., {Reynolds} J.~E.,  {Roy} A.~L., 1995, \mnras,
  276, 1373

\bibitem[\protect\citeauthoryear{{Seaquist} \& {Bell}}{{Seaquist} \&
  {Bell}}{1986}]{SeaquistBell1986}
{Seaquist} E.~R.,  {Bell} M.~B., 1986, \apjl, 303, L67

\bibitem[\protect\citeauthoryear{{Shaffer} \& {Marscher}}{{Shaffer} \&
  {Marscher}}{1979}]{ShafferMarscher1979}
{Shaffer} D.~B.,  {Marscher} A.~P., 1979, \apjl, 233, L105

\bibitem[\protect\citeauthoryear{{Simpson} et~al.}{{Simpson}
  et~al.}{1996}]{SimpsonMulchaeyWilson1996}
{Simpson} C., {Mulchaey} J.~S., {Wilson} A.~S., {Ward} M.~J.,  {Alonso-Herrero}
  A., 1996, \apjl, 457, L19

\bibitem[\protect\citeauthoryear{{Trotter} et~al.}{{Trotter}
  et~al.}{1998}]{TrotterGreenhillMoran1998}
{Trotter} A.~S., {Greenhill} L.~J., {Moran} J.~M., et~al., 1998, \apj, 495, 740

\bibitem[\protect\citeauthoryear{{Ulvestad} et~al.}{{Ulvestad}
  et~al.}{1999}]{UlvestadWrobelRoy1999}
{Ulvestad} J.~S., {Wrobel} J.~M., {Roy} A.~L., et~al., 1999, \apjl, 517, L81

\bibitem[\protect\citeauthoryear{{Warwick} et~al.}{{Warwick}
  et~al.}{1989}]{WarwickKoyamaInoue1989}
{Warwick} R.~S., {Koyama} K., {Inoue} H., et~al., 1989, \pasj, 41, 739

\bibitem[\protect\citeauthoryear{{Whiteoak} \& {Gardner}}{{Whiteoak} \&
  {Gardner}}{1987}]{WhiteoakGardner1987}
{Whiteoak} J.~B.,  {Gardner} F.~F., 1987, Proceedings of the Astronomical
  Society of Australia, 7, 88

\bibitem[\protect\citeauthoryear{{Wilson} et~al.}{{Wilson}
  et~al.}{1998}]{WilsonRoyUlvestad1998}
{Wilson} A.~S., {Roy} A.~L., {Ulvestad} J.~S., et~al., 1998, \apj, 505, 587

\end{thebibliography}
\end{document}